\documentclass[nonacm,sigplan,10pt]{acmart}
\renewcommand\footnotetextcopyrightpermission[1]{}
\usepackage[inline]{enumitem}
\usepackage[utf8]{inputenc}

\usepackage[]{hyperref}
\newcommand{\arch}[1]{LithOS}
\newcommand{\cmu}[1]{UniX}
\newcommand{\meta}[1]{Meta}

\newcommand{\us}{$\mu$s}
\newcommand{\ms}{ms}
\newcommand{\x}{$\times$}

\usepackage{makecell}
\usepackage{tikz}
\usepackage{siunitx}
\newcommand*\circled[1]{\raisebox{.5pt}{\textcircled{\raisebox{-.9pt} {#1}}}}

\usepackage[format=plain,labelfont=bf,textfont=it]{caption}

\usepackage[htt]{hyphenat}

\usepackage{algorithm}
\usepackage{algpseudocode}
\usepackage{color}
\usepackage{listings}
\usepackage{graphicx,multirow}
\usepackage[]{mdframed}

\definecolor{GrayCodeBlock}{RGB}{241,241,241}
\definecolor{BlackText}{RGB}{110,107,94}
\definecolor{RedTypename}{RGB}{220,86,17}
\definecolor{GreenString}{RGB}{96,172,57}
\definecolor{PurpleKeyword}{RGB}{184,84,212}
\definecolor{GrayComment}{RGB}{170,170,170}
\definecolor{GoldDocumentation}{RGB}{180,165,45}
\lstdefinelanguage{rust}
{
    columns=fullflexible,
    keepspaces=true,
    frame=single,
    framesep=0pt,
    framerule=0pt,
    framexleftmargin=4pt,
    framexrightmargin=4pt,
    framextopmargin=5pt,
    framexbottommargin=3pt,
    xleftmargin=4pt,
    xrightmargin=4pt,
    basicstyle=\ttfamily\scriptsize,
    keywords={
        true,false,
        unsafe,async,await,move,
        use,pub,crate,super,self,mod,
        struct,enum,fn,const,static,let,mut,ref,type,impl,dyn,trait,where,as,
        break,continue,if,else,while,for,loop,match,return,yield,in,
        sizeof, kernel, define,
    },
    keywordstyle=\color{blue},
    ndkeywords={
        bool,u8,u16,u32,u64,u128,i8,i16,i32,i64,i128,char,str,f32,
        Self,Option,Some,None,Result,Ok,Err,String,Box,Vec,Rc,Arc,Cell,RefCell,HashMap,BTreeMap, blockIdx, gridDim, 
        macro_rules
    },
    ndkeywordstyle=\color{RedTypename},
    comment=[l][\color{GreenString}\slshape]{//},
    morecomment=[s][\color{GrayComment}\slshape]{/*}{*/},
    morecomment=[l][\color{GoldDocumentation}\slshape]{///},
    morecomment=[s][\color{GoldDocumentation}\slshape]{/*!}{*/},
    morecomment=[l][\color{GoldDocumentation}\slshape]{//!},
    morecomment=[s][\color{RedTypename}]{\#![}{]},
    morecomment=[s][\color{RedTypename}]{\#[}{]},
    stringstyle=\color{GreenString},
    string=[b]",
    numbers=left,
    stepnumber=1,
}

\usepackage[export]{adjustbox}

\newcommand{\mypar}[1]{{\bf {#1}}}
\newcommand{\infTrainTailMPS}{4.7}

\newcommand{\infTrainTailTGS}{1.18}

\newcommand{\infTrainTptTGS}{1.35}

\newcommand{\rightsizingAVG}{26}
\newcommand{\rightsizingPerfAVG}{4}

\copyrightyear{2025}
\setcopyright{iw3c2w3g}

\newcommand\blfootnote[1]{%
  \begingroup
  \renewcommand\thefootnote{}\footnote{#1}%
  \addtocounter{footnote}{-1}%
  \endgroup
}

\begin{document}

\title[\arch{}: An Operating System for Efficient Machine Learning on GPUs]{\arch{}: An Operating System\\for Efficient Machine Learning on GPUs}

\author[P. H. Coppock, B. Zhang, E. H. Solomon, V. Kypriotis, L. Yang, B. Sharma, D. Schatzberg, T. C. Mowry, and D. Skarlatos]{
Patrick H. Coppock, Brian Zhang, Eliot H. Solomon, Vasilis Kypriotis, Leon Yang$^{\dag}$, Bikash Sharma$^{\dag}$,\\Dan Schatzberg$^{\dag}$, Todd C. Mowry, and Dimitrios Skarlatos
\\Carnegie Mellon University \quad $^{\dag}$Meta\\}

\begin{abstract}
The surging demand for GPUs in datacenters for machine learning (ML) workloads has made efficient GPU utilization crucial. However, meeting the diverse needs of individual ML models while optimizing resource usage is challenging. To enable transparent, fine-grained management of GPU resources
that maximizes GPU utilization and energy efficiency while maintaining
strong isolation, an operating systems (OS) approach is needed.  
Hence this paper introduces \arch{}, a first step towards a GPU OS. 

\arch{} includes the following new abstractions and mechanisms for efficient GPU resource management: (i) a novel \textit{TPC Scheduler} that supports spatial scheduling at the granularity of individual TPCs, unlocking efficient TPC stealing between workloads; (ii) transparent \textit{kernel atomization} to reduce head-of-line blocking and allow dynamic resource reallocation mid-execution; (iii) a lightweight \textit{hardware right-sizing} mechanism that dynamically determines the minimal TPC resources needed per atom; and (iv) a transparent \textit{power management} mechanism that reduces power consumption based upon in-flight work characteristics.

We implement \arch{} in Rust and evaluate its performance across a broad set of deep learning environments, comparing it to state-of-the-art solutions from NVIDIA and prior research.
For inference stacking, \arch{} reduces tail latencies by 13\x{} compared to MPS; compared to the best-performing SotA, it reduces tail latencies by 3\x{} while improving aggregate throughput by 1.6\x{}. 
Furthermore, in hybrid inference-training stacking, \arch{} reduces tail latencies by \infTrainTailMPS{}$\times$ compared to MPS; compared to the best-performing SotA, it reduces tail latencies by \infTrainTailTGS{}$\times$ while improving aggregate throughput by \infTrainTptTGS{}$\times$. 
Finally, for a modest performance hit under 4\%, \arch{}'s hardware right-sizing provides a quarter of GPU capacity savings on average, while for a 7\% hit, \arch{}'s transparent power management delivers a quarter of a GPU total energy savings on average.
Overall, \arch{} transparently increases GPU efficiency, establishing a foundation for future OS research on GPUs.
\end{abstract}

\maketitle

\section{Introduction}
\label{sec:intro}

\begin{figure}[t]
\centering
\includegraphics[width=0.85\linewidth]{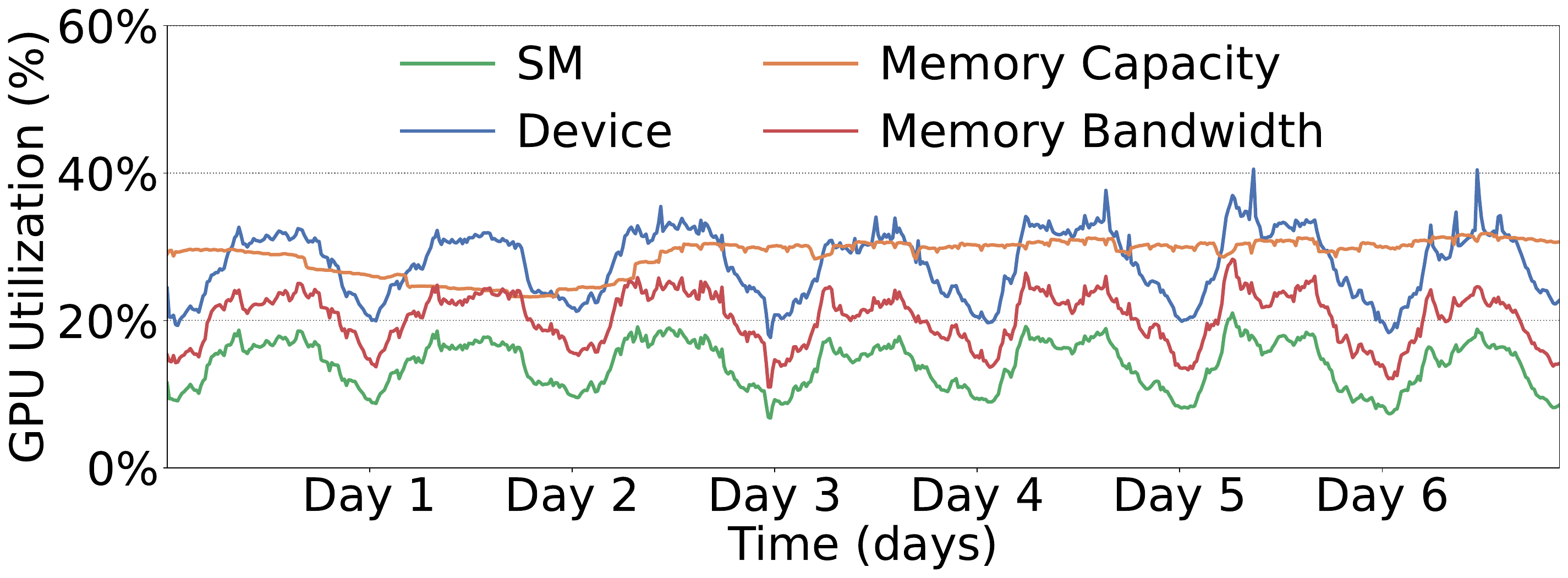}
\vspace{-4mm}
\caption{GPU utilization metrics over a week in a production Ads inference service at \meta{}.}
\vspace{-5mm}
\label{fig:gpu_util}
\end{figure}

The widespread adoption of machine learning (ML) workloads has led to massive GPU deployments across datacenters. However, despite growing concerns around power consumption and hardware supply constraints, GPU resources remain significantly underutilized. Public reports from Microsoft and Alibaba cite average and median GPU utilization rates of just 52\%~\cite{msrUtilization} and 10\%~\cite{antman}, respectively. Our analysis of a production \textit{Ads} service at \meta{} reveals similarly low utilization, averaging just 27\%, as shown in Figure~\ref{fig:gpu_util}. Given the high monetary cost and rising power demands---now exceeding 1,000W per GPU~\cite{h100Power,b100Power}---this is unsustainable.
\blfootnote{\copyright 2025. 
This work is licensed under the Creative Commons
Attribution-NonCommercial-NoDerivs~4.0 International
(\href{https://creativecommons.org/licenses/by-nc-nd/4.0/}{CC BY-NC-ND 4.0}) license.
Authors reserve their rights to
disseminate the work on their personal and corporate Web sites with
the appropriate attribution.}

It is challenging to achieve high utilization without GPU sharing. While dedicating a GPU to a single workload leads to high performance, individual workloads often fail to keep the GPU fully utilized: GPU cores idle on communication stalls, low batch sizes result in insufficient parallelism, dynamic request loads lead to overprovisioning, and so on~\cite{msft-low-gpu-util,reef,antman}. 
As GPUs become more powerful with increasing streaming multiprocessor (SM) counts and memory bandwidth~\cite{h100,b100Power}, achieving high utilization will become more challenging. 

One potential approach to GPU sharing is collocating \textit{latency-critical} (LC) tasks for which performance is of utmost importance with \textit{best-effort} (BE) tasks that lack hard deadlines.
However, existing systems do not offer a practical solution for prioritizing LC tasks over BE tasks when they contend for resources. Many approaches lack transparency, rendering them incompatible with large parts of the ML software stack~\cite{clockwork,TensorFlow-Serving,Clipper,INFaaS,Nexus,orion,abacus,Muxflow,reef,fgpu,pipeswitch,paella}. For instance, some are tied to specific versions of frameworks like PyTorch or TVM that are no longer maintained~\cite{antman,orion,clockwork,pipeswitch,abacus}. Other solutions like TGS~\cite{tgs} or Clockwork~\cite{clockwork} fall short of achieving high GPU utilization due to limited temporal scheduling that cannot execute multiple models in parallel.
Spatial scheduling solutions, including NVIDIA's MPS~\cite{mps} and MIG~\cite{mig} or research efforts like REEF~\cite{reef}, Orion~\cite{orion}, and others~\cite{abacus}, enable parallel execution of multiple applications. However, they are too coarse-grained, scheduling entire inference requests, training batches, or DNN operators, resulting in low utilization and head-of-line (HoL) blocking~\cite{reef,abacus, tgs,orion,pipeswitch,Muxflow,INFaaS,gpulets,MIG-serving,GSLICE,miso,iGniter,paella}. Efficient multitenant scheduling on GPUs has remained elusive.

Beyond collocation mechanisms, to address GPU inefficiencies without sacrificing performance or transparency, datacenter GPU management must evolve past static provisioning. Current systems fail to account for changing deep learning workload characteristics---such as fluctuating levels of compute intensity and parallelism across different models and execution phases. This is another reason why GPUs are often not efficiently utilized, even as they draw significant power. Bridging this gap requires new approaches that can adapt resource allocation and power consumption to the fine-grained characteristics of ML workloads.

This utilization crisis is in stark contrast with the situation for CPUs, where time-sharing operating systems allocate tasks to cores via inexpensive context switches, providing isolation, resource allocation, power management, and transparency. 
The extreme data-parallel nature of GPUs imposes different trade-offs than do CPUs, but also exposes the limitations of current abstractions built around compilers, frameworks, and drivers. To transparently improve utilization and efficiency, we believe that GPUs must evolve toward an operating system model---one that brings first-class support for control, isolation, and resource management.
\blfootnote{Contact: \href{mailto:dskarlat@cs.cmu.edu}{dskarlat@cs.cmu.edu} }

\subsection{Our Approach: An Operating System for GPUs}

To address datacenter GPU efficiency challenges, we introduce \arch{}, which brings an operating system approach to deep learning on GPUs.
\arch{} is fully transparent to the ML stack, allowing seamless integration without requiring any modifications to models, runtimes, or frameworks.
\arch{} moves the bulk of GPU scheduling from proprietary drivers and hardware into software, allowing, for the first time, fine-grained temporal and spatial scheduling of ML workloads. \arch{} operates at the granularity of individual kernel thread blocks that are dynamically mapped onto the GPU's thread processing clusters (TPCs). To achieve this, \arch{} introduces novel abstractions and mechanisms that decouple kernel work submission from thread block execution on GPUs, enabling intelligent scheduling decisions, resource allocation, and power management.

First, \arch{} introduces a novel fine-grained \textit{TPC Scheduler} that asynchronously determines the compute unit allocation and submission time for each piece of work. It enables precise control at the granularity of individual TPCs, providing strong isolation between workloads. The scheduler is guided toward efficient scheduling decisions by an online kernel latency predictor and incorporates a technique called \textit{TPC Stealing} to improve GPU utilization.

To overcome the lack of hardware preemption, \arch{} introduces a \textit{kernel atomization} mechanism that transparently---without compiler, runtime, source, or PTX code modifications---splits kernels into independently schedulable units called \textit{atoms}. Each atom consists of a subset of a kernel’s thread blocks, reducing head-of-line blocking and cross-workload interference. Atomization also enables \arch{} to dynamically reconfigure TPC allocations mid-execution, allowing scheduling flexibility that is impossible with monolithic kernels.

Building on this foundation, \arch{} introduces a dynamic \textit{hardware right-sizing} mechanism that uses lightweight modeling to determine the minimal TPC resources required for each kernel and its atoms, yielding significant capacity savings. Finally, \arch{} presents a fine-grained \textit{power management} mechanism that adjusts the GPU's frequency in response to the characteristics of in-flight work, achieving substantial energy savings.

We implement \arch{} in Rust and evaluate its performance across a broad set of deep learning environments, comparing it to state-of-the-art solutions from NVIDIA and prior research.
For inference stacking, \arch{} reduces tail latencies by 13\x{} compared to MPS; compared to the best-performing SotA, it reduces tail latencies by 3\x{} while improving aggregate throughput by 1.6\x{}. 
Furthermore, in hybrid inference-training stacking, \arch{} reduces tail latencies by \infTrainTailMPS{}$\times$ compared to MPS; compared to the best-performing SotA, it reduces tail latencies by \infTrainTailTGS{}$\times$ while improving aggregate throughput by \infTrainTptTGS{}$\times$. 
Finally, for a modest performance hit under 4\%, \arch{}'s hardware right-sizing provides a quarter of GPU capacity savings on average, while for a 7\% hit, \arch{}'s transparent power management delivers a quarter of a GPU total energy savings on average.

Overall, \arch{} transparently increases GPU efficiency, establishing a foundation for future OS research on GPUs.

This paper makes the following contributions:
\begin{itemize}[leftmargin=10pt,parsep=1pt,topsep=2pt]
    
    \item A comprehensive study of an \textit{Ads} inference service at \meta{}, highlighting the behavior of production ML models and the challenges of GPU underutilization. 
    \item A fine-grained spatial \textit{TPC Scheduler} that dynamically allocates TPCs using \textit{TPC Stealing} to boost utilization.
    \item A transparent \textit{Kernel Atomizer} that independently schedules subsets of kernel thread blocks, unlocking efficiency.
    \item A dynamic \textit{hardware right-sizing} mechanism that optimizes TPC allocations for significant capacity savings.
    \item A transparent \textit{power management} mechanism that adjusts frequency based on kernel characteristics to save energy.
    \item The design of \arch{}, a step towards an OS for GPUs. 
    \item The evaluation of \arch{} across varied ML environments.
\end{itemize}
\section{Background and Related Work}
\label{sec:background}
In this section, we first present a brief background on NVIDIA GPU architectures and then cover related work.

\subsection{A Brief Background on GPUs}

\begin{figure}[t]
\centering
\includegraphics[width=0.6\columnwidth]{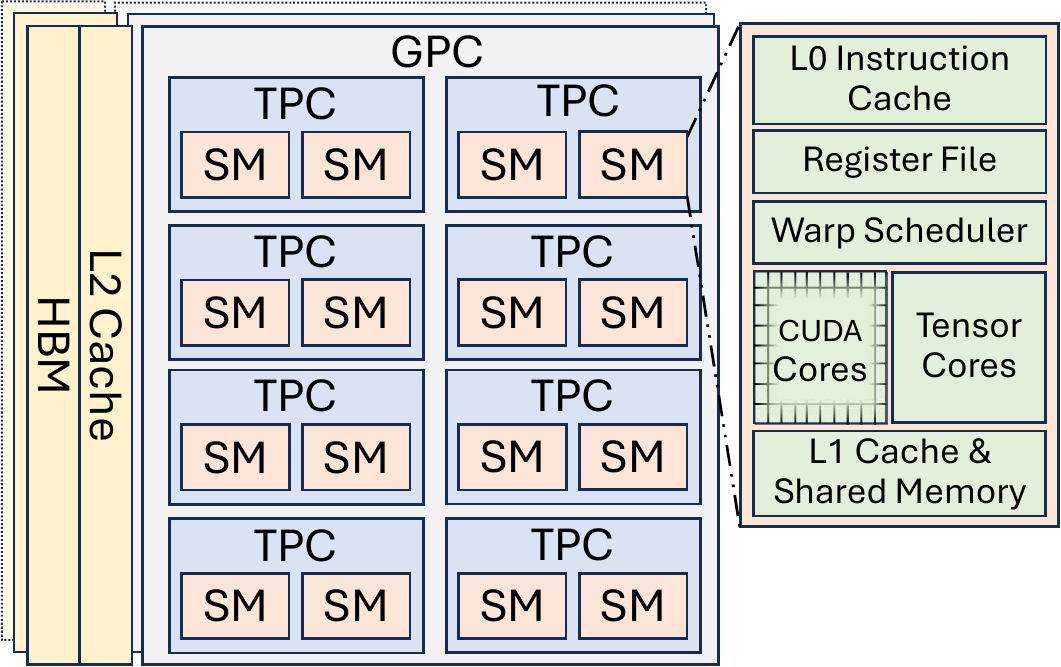}
\vspace{-3mm}
\caption{GPU Architecture.}
\vspace{-4mm}
\label{fig:gpu_arch}
\Description{}{}
\end{figure}

\mypar{GPU Architecture.} Modern GPUs have immense hardware resources catering to the needs of ML workloads. 
Figure~\ref{fig:gpu_arch} depicts a typical GPU architecture. Each GPU consists of several General Processing Clusters (GPCs). Each GPC is a collection of multiple Thread Processing Clusters (TPCs) and each TPC includes a small number of Streaming Multiprocessors (SMs). Each SM is composed of tens of cores. For example, NVIDIA's H100~\cite{h100} includes 8 GPCs, 9 TPCs per GPC, 2 SMs per TPC, and 128 cores per SM. 

\mypar{GPU Programming.}
GPU applications are composed of kernels that execute specific operators (e.g., convolution). A kernel defines its resources---thread blocks, threads, registers, and shared memory---at launch time. Programmers divide a kernel's work among the thread blocks. Each thread block executes on an SM and consists of multiple SIMD threads.

\mypar{GPU Streams.} CUDA streams enable concurrent execution of independent tasks, similar to CPU threads. Stream work is executed in FIFO order. Some CUDA calls are asynchronous, while others wait for all previous tasks to finish.

\subsection{Related Work}
\label{sec:related}

\mypar{Cooperative multitenancy.} Cooperative scheduling involves tenants coordinating to share resources, typically at the ML framework level, with all models running in the same process~\cite{clockwork,TensorFlow-Serving,Clipper,INFaaS,Nexus,orion,abacus,Muxflow,reef,fgpu,pipeswitch,paella}. These approaches require custom ML frameworks and are hence limited by their inability to support arbitrary applications. Some also rely on extensive offline profiling \cite{orion, reef} or kernel source modifications \cite{paella, reef} which are impractical at scale. Finally, any non-cooperating tenant invalidates guarantees made by the runtime, making adoption difficult in practice.

\mypar{Transparent multitenancy.} Transparent GPU sharing solutions support unmodified applications. They include native mechanisms like time slicing, MPS~\cite{mps}, and MIG~\cite{mig} offered by NVIDIA. Nearly all prior software solutions are not transparent and rely on application or framework modifications. TGS~\cite{tgs} is one exception that offers transparent sharing between containerized applications. In practice, uncooperative tasks and limited application-specific information make transparent multitasking a serious challenge.  

\mypar{Temporal multitenancy.} Temporal multitenancy dedicates the entire GPU to a single task at a time via native time slicing or software scheduling. Some approaches work at the level of entire inference requests (e.g., Clipper~\cite{Clipper}, Nexus~\cite{Nexus}, TensorFlow-Serving~\cite{TensorFlow-Serving}, Clockwork~\cite{clockwork}, and INFaaS~\cite{INFaaS}), while others schedule individual GPU kernels (e.g., PipeSwitch~\cite{pipeswitch}, AntMan~\cite{antman}, and TGS~\cite{tgs}).
\textit{Time slicing} is NVIDIA's default temporal multitenancy solution. It shares the GPU in a round-robin fashion, giving each task exclusive access for several milliseconds.
These methods execute only one job at a time, leading to low utilization.

\mypar{Spatial multitenancy.} Spatial multitenancy typically builds on MIG or MPS to enable multiple applications to run concurrently on a GPU and improve utilization.
\textit{MPS} multiplexes multiple GPU contexts onto one, allowing multiple tasks to use the GPU concurrently. This can yield greater throughput but leads to performance interference.
\textit{MIG} partitions the GPU's compute and memory resources along GPC boundaries, providing strong hardware isolation. However, the coarse granularity of its partitioning and steep reconfiguration overheads (>5s~\cite{mig-reconfiguration-is-5s}) can leave resources idle. 

\begin{figure}[t]
\centering
\includegraphics[width=\columnwidth]{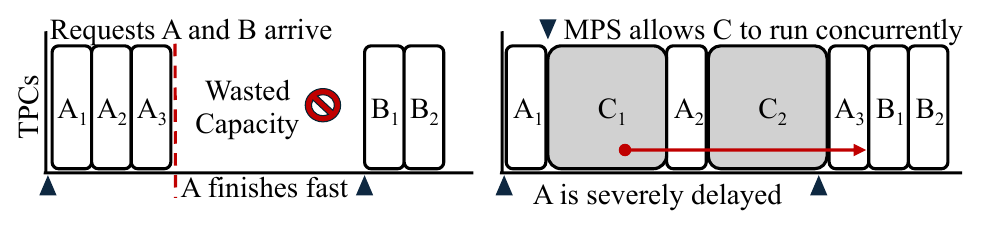}
\vspace{-8mm}
\caption{GPU timeline showcasing the pitfalls of MPS.}
\vspace{-4mm}
\label{fig:native-timeline}
\Description{}{}
\end{figure}

Like temporal systems, existing spatial sharing systems are coarse-grained, operating at the level of inference requests or kernels. Their goal is to protect latency-critical (LC) applications by restricting kernels launched by other jobs~\cite{abacus,orion} or limiting GPU resources allocated to best-effort tasks, as seen in systems like REEF~\cite{reef}, MuxFlow~\cite{Muxflow}, and others~\cite{fgpu,gpulets,MIG-serving,GSLICE,miso,iGniter,paella}. However, the coarseness of these approaches limits control over GPU resources, often leading to HoL blocking, low utilization, and interference. 
Figure~\ref{fig:native-timeline} highlights the challenges of spatial sharing. In Figure~\ref{fig:native-timeline}(a), a single workload runs on the GPU, issuing two requests with five total kernels. This results in fast kernel completion for $A$ and $B$ but leaves much of the GPU underutilized. When MPS enables concurrent execution of multiple tasks in Figure~\ref{fig:native-timeline}(b), utilization is improved, but the original task's requests face significant delays.
Overall, prior works have tackled some multitenant ML scheduling challenges but fail to offer a complete, transparent solution. Importantly, prior temporal and spatial strategies operate at a coarse granularity, limit utilization, and cause HoL blocking, which interferes with collocated workloads.

\mypar{Right-sizing.} Prior efforts have explored GPU job right-sizing to improve resource efficiency. However, these approaches often rely on hardware modifications~\cite{HSM,KRISP}, lack transparency to application software~\cite{GSLICE,smCompactor,gpulets,PARIS-ELSA,SGDRC}, and depend on offline profiling~\cite{GSLICE,gpulets,PARIS-ELSA,KRISP,CoFRIS,SGDRC}. Crucially, most existing solutions operate at the granularity of entire jobs, which limits their ability to fully exploit the benefits of fine-grained right-sizing and can lead to suboptimal performance.

\mypar{Dynamic Voltage Frequency Scaling.} Recent efforts~\cite{stojkovic2024, kakolyris2024,ZhangDVFS, QiuDVFS,POLCA} have applied dynamic voltage frequency scaling (DVFS) to minimize the power consumption of GPUs with a particular focus on LLM inference clusters~\cite{stojkovic2024, kakolyris2024}. Such approaches are based on extensive offline profiling across several input lengths and train dedicated output length predictors, failing to provide a transparent mechanism. Prior work on DVFS operates at a coarser granularity, observing the performance of the whole inference request and missing finer optimization opportunities.

\section{Motivation}
\label{sec:motivation}
In this section, we showcase a detailed study of production GPU infrastructure challenges and opportunities.

\subsection{Understanding GPU Utilization in Datacenters}

To understand GPU utilization in datacenters, we analyze a subset of \textit{Ads} inference services at \meta{}, which serve deep learning models across its fleet. At \meta{}, \textit{Ads} inference relies in part on NVIDIA H100 GPU nodes. Each node has 8 GPUs, each partitioned via MIG. The production service performs offline analysis of each model, assigning models to GPU/MIG partitions for deployment. The goal is to meet tight SLAs on tail response times requirements for each model.

\mypar{GPU Utilization.} In Figure~\ref{fig:gpu_util}, we show GPU compute and memory utilization over a week. Device compute utilization ranges from 17\% to 40\%, averaging 27\%. SM utilization is even lower, averaging 14\%, with peaks at 21\% and lows of 6.7\%. Memory bandwidth utilization averages 20\%. Overall, utilization follows a diurnal pattern tied to inference traffic. However, memory capacity remains steady at 28\%, as models are kept loaded in GPU memory to meet tight SLAs. These SLAs also enforce small batch sizes, preventing full GPU resource saturation even at high request loads.

\begin{figure}[t]
\centering
\includegraphics[width=0.8\columnwidth]{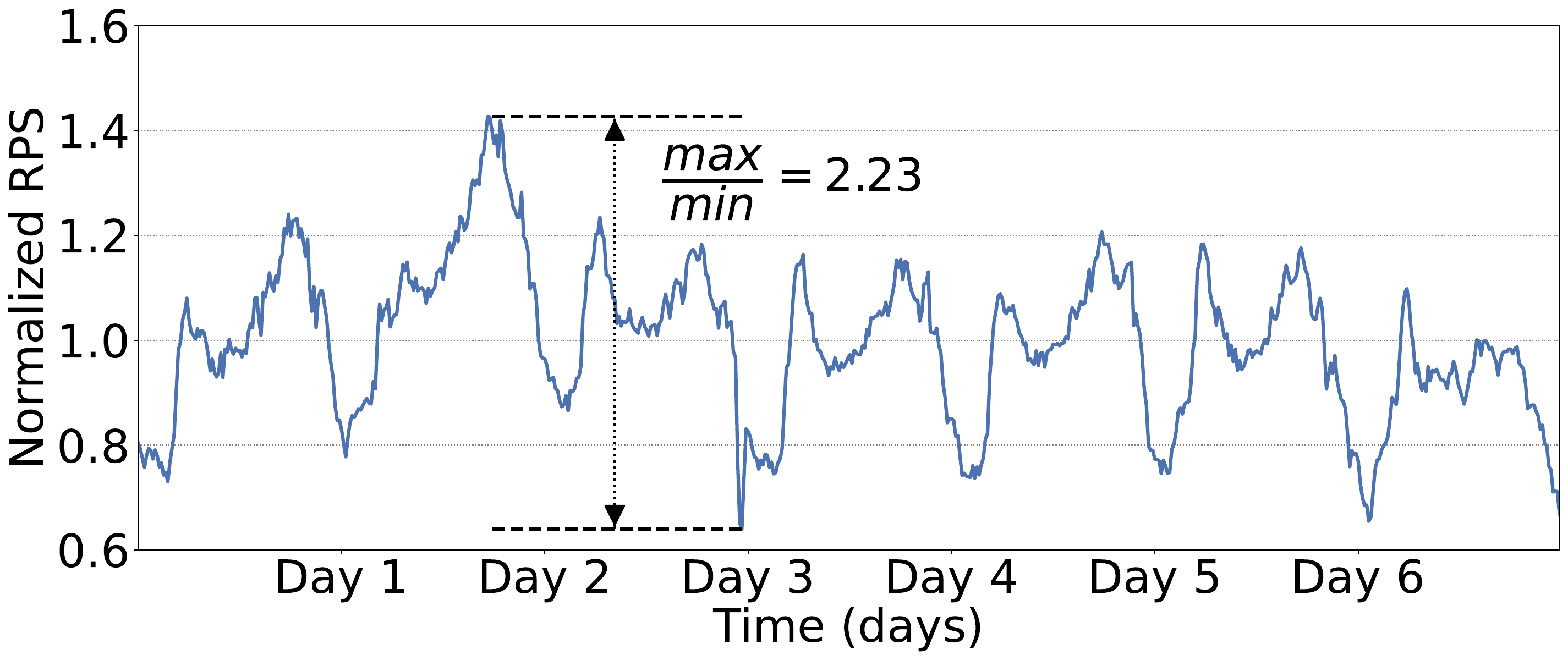}
\vspace{-4mm}
\caption{Mean normalized traffic.} 
\vspace{-2mm}
\label{fig:rps}
\Description{}{}
\end{figure}

\begin{figure}[t]
\centering
\includegraphics[width=0.8\columnwidth]{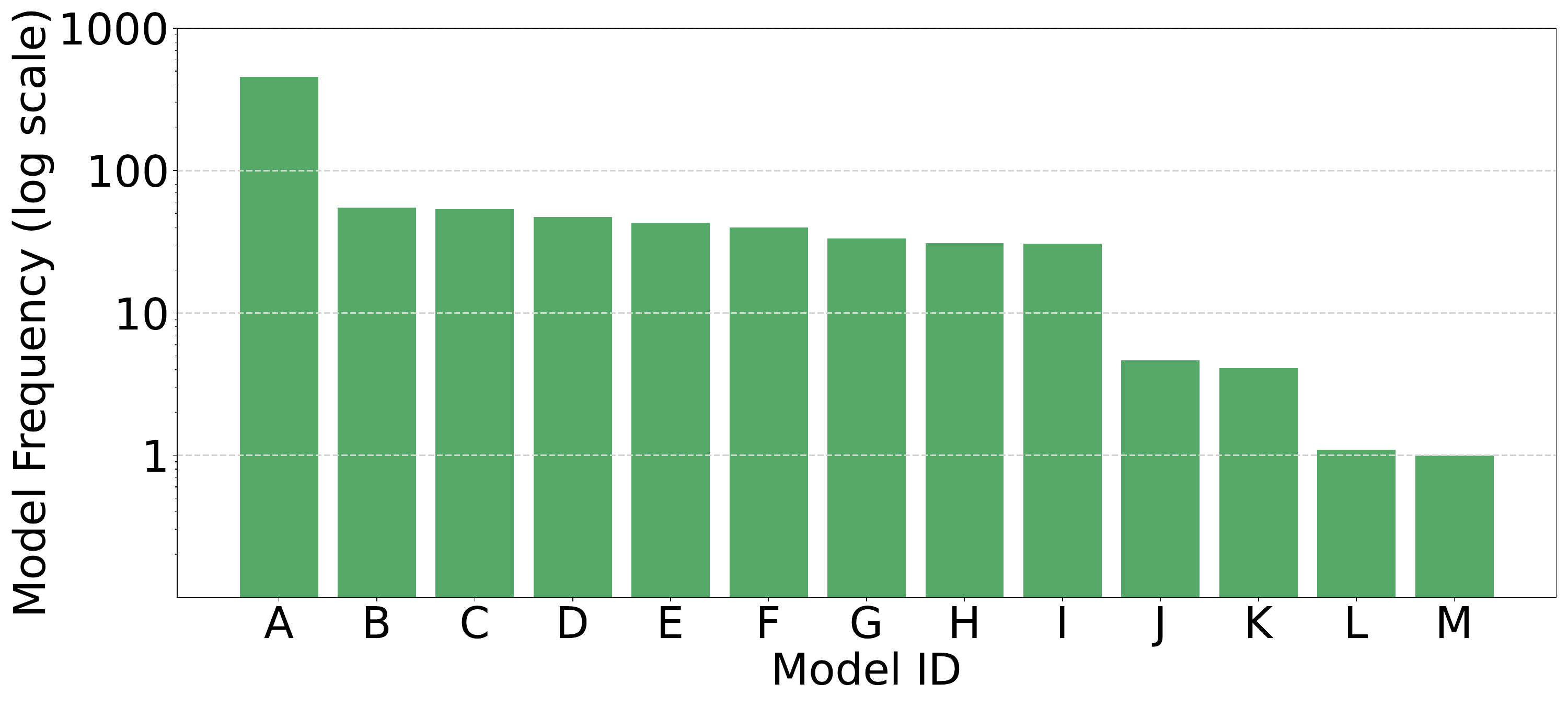}
\vspace{-4mm}
\caption{Model frequency distribution.}
\vspace{-2mm}
\label{fig:model_frequency}
\Description{}{}
\end{figure}

\mypar{Inference Traffic.} To investigate low GPU utilization, we first examine inference traffic. Figure~\ref{fig:rps} shows the mean-normalized requests per second (RPS) over a week, revealing a diurnal pattern. RPS can scale by 2.2$\times$ between minimum and maximum traffic, closely correlating with the GPU utilization trends shown in Figure~\ref{fig:gpu_util}.
Next, we analyze model request frequencies. We sample thirteen of the most commonly used models and plot in Figure~\ref{fig:model_frequency} the normalized frequency of inference requests over the same week. The distribution's variance is significant, with the most popular model \textit{A} receiving several hundred times more requests than the least popular model \textit{M}. Over-provisioning GPUs for such a wide request distribution can lead to underutilization, particularly for less popular models.

\begin{figure}[t]
\centering
\includegraphics[width=0.8\columnwidth]{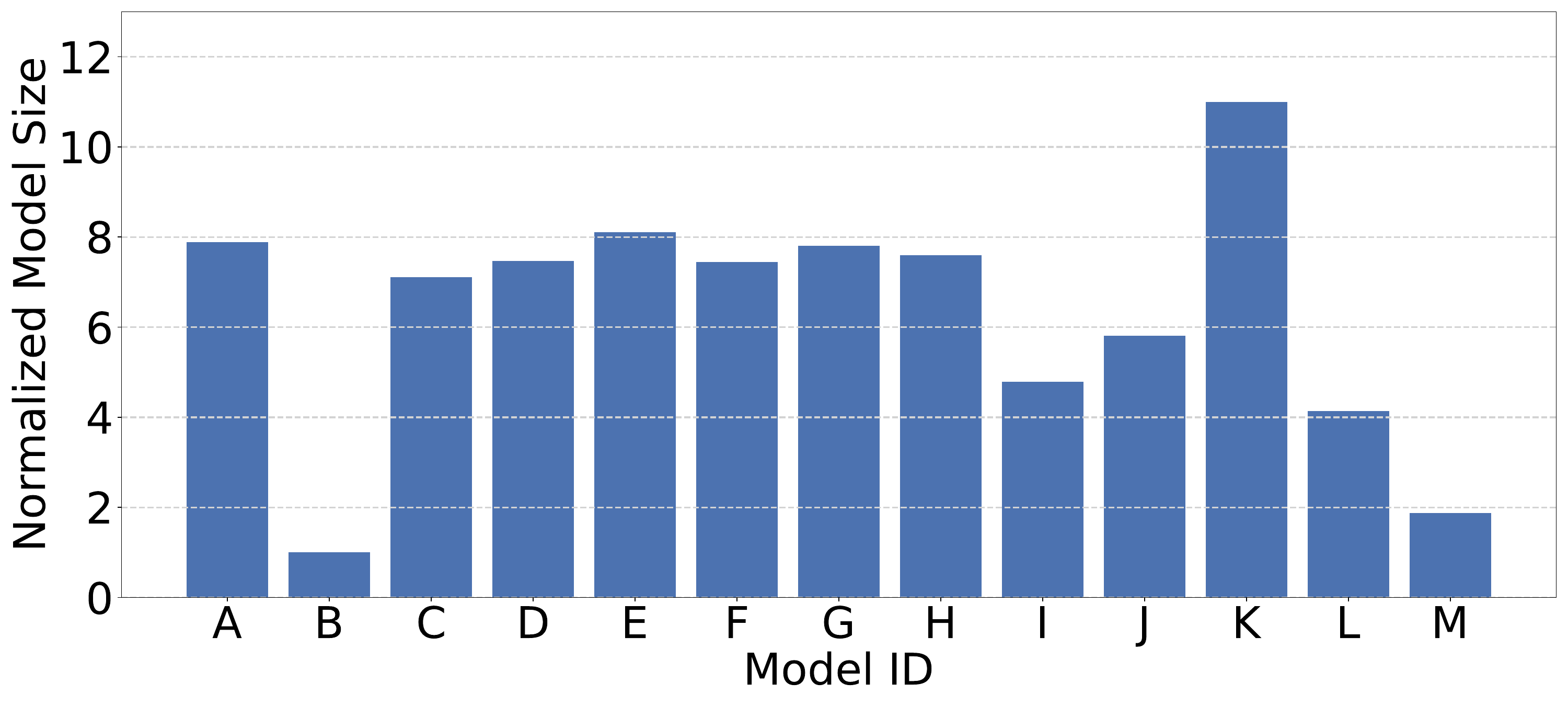}
\vspace{-4mm}
\caption{Model size distribution.}
\vspace{-2mm}
\label{fig:model_size}
\Description{}{}
\end{figure}

\mypar{Model Sizes.}
To better understand GPU utilization, we examine the sizes of the most commonly used models based on weights, parameters, and embeddings. As shown in Figure~\ref{fig:model_size}, model sizes vary significantly, with a more than a 10$\times$ difference between the largest and smallest models. Half are relatively large, while the rest are smaller. Both large and small models are frequently used: for example, the smallest model \textit{B} has usage comparable to larger models \textit{E} and \textit{G}. This highlights the opportunity to collocate models of different sizes while meeting each of their service-level agreements.

\mypar{GPU Sharing Limitations and Takeaways.} 
Despite the urgent need to improve GPU utilization, datacenters often rely on limited GPU sharing or hardware approaches like MIG due to requirements for compatibility and transparency within the ML software stack. Non-transparent solutions that require framework or application changes for multitenancy are impractical at scale, given the complexity of maintaining multiple ML frameworks, runtimes, and compilers.
Importantly, in the rapidly evolving ML space, transparent solutions help avoid the risk of locking infrastructure into rigid, outdated designs.
Based on these insights, we design \arch{} as a fully transparent OS for efficient ML multitenancy. 

\section{\arch{} Design}
\label{sec:design}

We propose \arch{}, an OS designed to address GPU inefficiencies in datacenters. \arch{} operates transparently across the ML stack, enabling efficient machine learning on GPUs.

\begin{figure}[t]
\centering
\includegraphics[width=0.8\columnwidth]{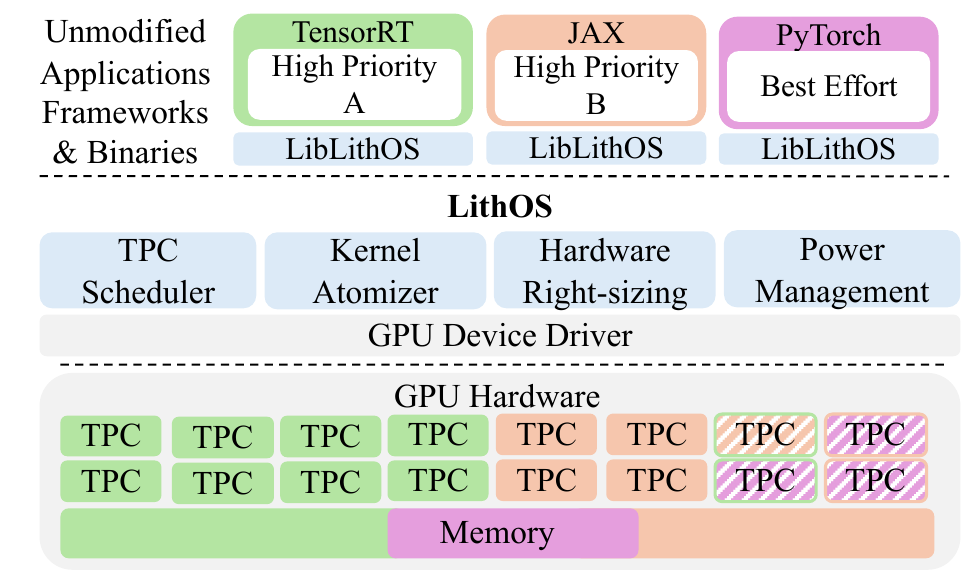}
\vspace{-4mm}
\caption{\arch{} architecture overview.}
\vspace{-4mm}
\label{fig:overview}
\Description{}{}
\end{figure}

\begin{figure*}[t]
\centering
\includegraphics[width=0.7\textwidth]{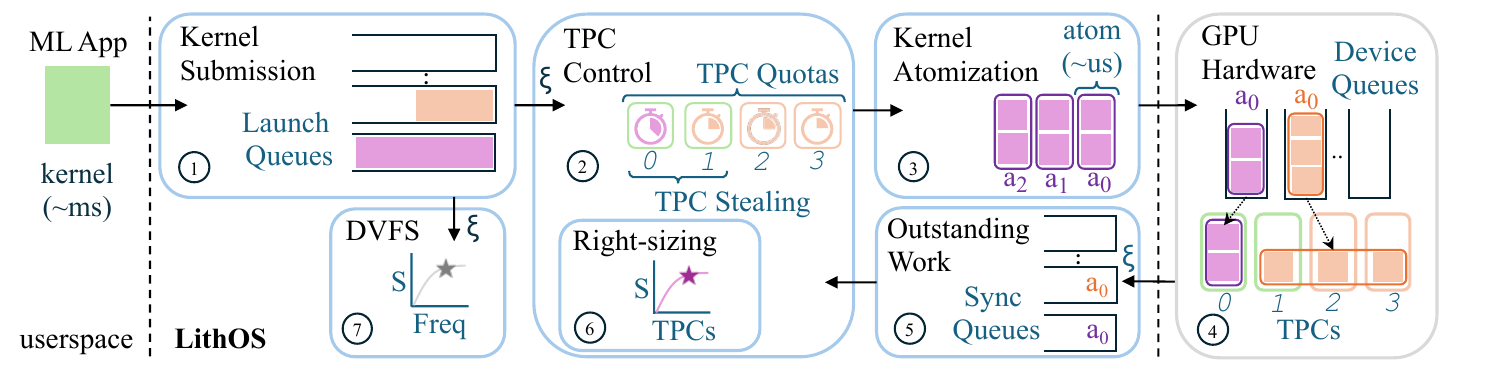}
\vspace{-5mm}
\caption{\arch{} operations overview.}
\vspace{-2mm}
\label{fig:op-timeline}
\Description{}{}
\end{figure*}

\subsection{Architecture Overview}

Figure~\ref{fig:overview} presents \arch{}'s architecture. \arch{} runs on CPU cores and interposes at the driver level, providing a dynamically linked library, Lib\arch{}, that mimics the native CUDA library. As a GPU operating system, \arch{} maintains a system-wide view of GPU state across applications with varying priorities, enabling efficient scheduling and management.
Applications follow the CUDA programming model and submit kernels to \arch{}, which decouples submission from GPU execution. This transparently shifts scheduling control from the driver and hardware to the \arch{} layer. The \textit{TPC Scheduler} manages resources at the granularity of individual TPCs, unlocking \textit{TPC Stealing} opportunities. Idle TPCs are lent to other tasks, improving utilization.

\arch{} also introduces the \textit{Kernel Atomizer}, which—without access to application source or PTX code—transparently breaks kernels into smaller thread block chunks called \textit{atoms}. This enables finer-grained GPU scheduling and reduces head-of-line (HoL) blocking.
Building on fine-grained control, \arch{} supports dynamic hardware right-sizing, using lightweight models to reduce TPC allocations for individual kernels and atoms, yielding substantial capacity savings. Finally, \arch{} applies transparent fine-grained DVFS, adjusting GPU frequency based on in-flight work to save energy.
Together, these mechanisms enable intelligent scheduling policies that maximize GPU utilization and efficiency across diverse ML workloads. The rest of this section details how these mechanisms operate and interact, referencing Figure~\ref{fig:op-timeline}.

\subsection{Interface with Userspace}
\label{sec:summary}

\mypar{Kernel Submission.}
Applications interact with \arch{} via \textit{launch queues} that buffer work (Figure~\ref{fig:op-timeline}, Step~\circled{1}), giving \arch{} full control over when work is dispatched to the GPU. This is important because once submitted, a kernel’s priority or resources cannot be changed, nor can it be rescheduled. Eagerly dispatching work can lead to sub-optimal scheduling. \arch{} therefore defers dispatch to minimize outstanding work on the GPU.
A launch queue is created when an application creates a stream via \texttt{cuStreamCreate}. On asynchronous CUDA calls like \texttt{cuLaunchKernel}, \arch{} enqueues the kernel and returns control to the application.

\mypar{Compute Quotas.} \arch{} allows users and system administrators to define GPU resource limits, exposing TPC quotas (Figure~\ref{fig:op-timeline}, Step~\circled{2}) that guarantee each application access to a specified number of TPCs when work is available. Internally, \arch{} manages TPCs analogously to how a traditional OS manages CPU cores, enabling fine-grained control over GPU resources.
As we will see next, \arch{} relies on a highly efficient \textit{TPC Scheduler} that interacts with launch queues and TPC quotas to optimize GPU utilization and efficiency. 

\subsection{TPC Scheduler}
\label{sec:scheduler}

\arch{} introduces a novel scheduler that operates at the granularity of individual TPCs, offering several advantages. TPC-level control enables fine-grained GPU resource management. Unlike static partitioning schemes like MIG, \arch{} supports dynamic, on-the-fly TPC allocation, allowing each kernel to run on a different set of TPCs without reconfiguration overhead. This flexibility maximizes utilization without coarse partitioning or slow reallocation.
Kernels are scheduled on their assigned TPCs, ensuring guaranteed resources for high-priority applications. However, as shown in Section~\ref{sec:motivation}, fixed allocations often leave TPCs idle due to traffic patterns or model variability. To address this, \arch{} employs dynamic scheduling and TPC Stealing to reassign idle resources. We believe TPC scheduling lays the foundation for evolving GPU policies, much like CPU scheduling has matured over time~\cite{258884, 227649}.

\mypar{Operation.} At a high level, the TPC Scheduler uses dispatcher threads to monitor launch queues (Figure~\ref{fig:op-timeline}, Step~\circled{1}) and submit work to the GPU. A key goal is to keep the GPU busy while maintaining scheduling flexibility. The scheduler faces two main challenges: varying kernel durations and balancing flexibility with GPU starvation. To address the former, it applies Kernel Atomization (Figure~\ref{fig:op-timeline}, Step~\circled{3}, Section~\ref{sec:atomizer}) to split long-running kernels into smaller thread block chunks called \textit{atoms}. To address the latter, it tracks outstanding work via sync queues (Figure~\ref{fig:op-timeline}, Step~\circled{5}), throttling submissions until the backlog drops below a tunable threshold. A dedicated Tracker thread monitors task completion and updates scheduler state.

\begin{figure*}[t]
\centering
\includegraphics[width=0.9\textwidth]{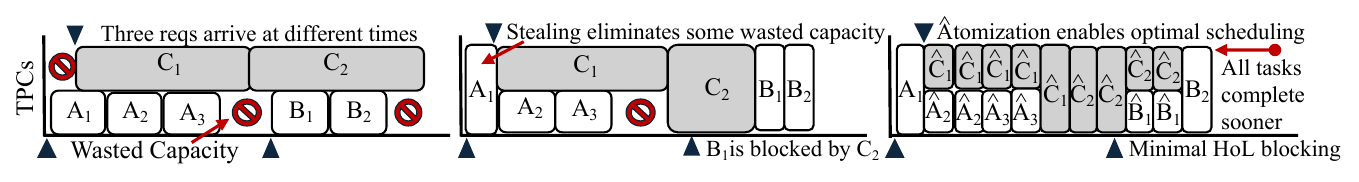}
\vspace{-5mm}
\caption{GPU timeline for two workloads showcasing (a) TPC Scheduling, (B) Stealing, and (C) Atomization.}
\vspace{-2mm}
\label{fig:timeline}
\Description{}{}
\end{figure*}

\mypar{TPC Stealing.} To improve work conservation, the scheduler dynamically reassigns underutilized TPCs across applications. In Figure~\ref{fig:timeline}(a), static allocation leads to idle TPCs. In Figure~\ref{fig:timeline}(b), stealing allows $A_1$ to borrow TPCs from an idle workload, reducing waste. However, this may cause head-of-line (HoL) blocking from priority inversion if a new request $B$ is delayed by $C_2$ occupying the stolen TPCs.
To mitigate this, the scheduler adopts a layered strategy. It maintains per-TPC timers informed by a latency prediction module, estimating kernel (and atom) durations at submission time. These timers help avoid stealing from long-running TPCs. As tasks complete, sync queues are cleared and timers updated—potentially refining predictions (Section~\ref{sec:learning}). \arch{} also limits outstanding atoms and uses lower hardware stream priorities for work on stolen TPCs. Combined with kernel atomization, these mechanisms boost utilization while minimizing interference.

\subsection{Kernel Atomizer}
\label{sec:atomizer}

At the core of \arch{} lies the \textit{Kernel Atomizer}. The Kernel Atomizer transforms kernels into small chunks called \textit{atoms}, each containing a subset of the grid's thread blocks (Figure~\ref{fig:op-timeline}, Step~\circled{3}). Importantly, the Kernel Atomizer operates without any access to source or PTX code, making it fully transparent to the entire ML software stack. This allows \arch{} to dispatch work at thread-block rather than kernel granularity. This is a critical requirement for an OS targeting GPUs, as the execution time of kernels can vary wildly from a few microseconds to tens of milliseconds.

\begin{figure}[t]
\centering
\includegraphics[width=1\columnwidth]{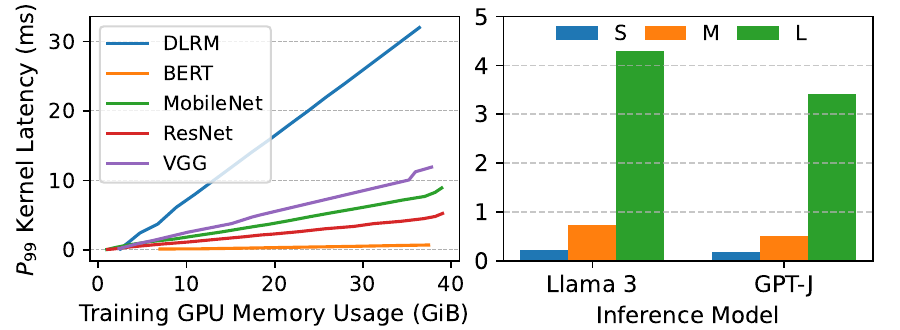}
\vspace{-6mm}
\caption{(a) $P_{99}$ kernel latency at different training batch sizes normalized to memory usage. (b) $P_{99}$ kernel latency for different inference prompt sequence lengths for LLMs. }
\vspace{-2mm}
\label{fig:latency_vs_bs}
\Description{}{}
\end{figure}

\mypar{Impact of Kernel Scheduling on Latency.} To illustrate the need for kernel atomization, Figure~\ref{fig:latency_vs_bs} presents $P_{99}$ kernel latencies across various training and inference workloads. Figure~\ref{fig:latency_vs_bs}(a) shows how $P_{99}$ latency increases with larger training batch sizes. Since the typical batch size for each model varies, we normalize by plotting memory usage at each size. Most models quickly produce long-running kernels lasting several milliseconds, with DLRM~\cite{naumov2019deeplearningrecommendationmodel} standing out with kernel latencies exceeding 30ms. While training workloads are the major culprit, in Figure~\ref{fig:latency_vs_bs}(b) we see that large language model (LLM) inference based on a trace from Microsoft Azure~\cite{stojkovic2024} containing small (\textit{S}), medium (\textit{M}), and large (\textit{L}) prompt lengths can also produce several-millisecond-long kernels for large prompts. Based on this analysis and given that models can have very tight SLO constraints (in the low tens of milliseconds), we guide the design of \arch{} toward a finer-grained scheduling unit that mitigates head-of-line blocking effects.

\mypar{Operation.} When a long-running kernel is about to be scheduled, \arch{} predicts the duration of the kernel given its TPC assignment using the predictor module (detailed in Section~\ref{sec:learning}). \arch{} then computes the number of atoms into which to split the kernel by dividing the predicted kernel duration by a tunable parameter called the \texttt{atom\_duration}. If this parameter is set too low, an atomized kernel may actually take longer to complete. Crucially, \arch{} is able to transparently chunk kernels into atoms at runtime. Atoms are then submitted to the GPU and can be scheduled on the TPCs dictated by the TPC Scheduler (Figure~\ref{fig:op-timeline}, \circled{4}). As a result, \arch{} resolves a major challenge faced by prior works that operate higher in the stack: the Kernel Atomizer works on applications written in any framework, that use any libraries (including closed-source ones like cuDNN), and are built with any compiler. 

To understand the benefits of scheduling at atom granularity, we return to Figure~\ref{fig:timeline}(b). Stealing improves the schedule but does not eliminate HoL blocking and wasted capacity. By dividing the kernels into atoms, work can be packed more tightly, as in Figure~\ref{fig:timeline}(c), and TPC allocations can be dynamically adjusted throughout a kernel's execution. Now, $B_1$ is no longer blocked by $C_2$, as stealing is disabled for the latter's subsequent atoms $\hat{C}_2$ once request $B$ is submitted.

To demonstrate how \arch{}’s Kernel Atomizer operates, we consider a \texttt{Conv} kernel with a grid dimension of \{8,8,1\}, resulting in 64 blocks with \texttt{block\_idx} ranging from 0 to 63. Instead of launching the \texttt{Conv} kernel directly, \arch{} invokes a \texttt{Prelude} kernel, which calls into the original kernel using the same launch configuration. The prelude kernel is shown in Algorithm~\ref{alg:prelude}.
At a high level, it checks whether \texttt{block\_idx} falls within a specified range—calling \texttt{Conv} if so, or exiting early otherwise. For example, to partition the grid into 2 atoms, the kernel atomizer launches the prelude twice with block index ranges [0,32) and [32,64).
Using this technique, \arch{} can divide the kernel into up to 64 atoms. By specifying non-overlapping block ranges, the atomizer ensures each block is executed once, maintaining correctness.

\begin{algorithm}[t]
\caption{Prelude Kernel Pseudocode.}\label{alg:prelude}
\begin{lstlisting}[language=rust]
kernel fn prelude(*args):
    let atom : *const AtomMetadata = AtomMetadataAddr as _
    let block_idx = blockIdx.z * gridDim.y * gridDim.x
                  + blockIdx.y * gridDim.x
                  + blockIdx.x
    if atom->block_idx_lo <= block_idx < atom->block_idx_hi:
        atom->kernel_entrypoint(*args)
\end{lstlisting}
\vspace{-3mm}
\end{algorithm}

\mypar{Atomization Considerations.} Kernels launch with an explicit set of resources; thus, the kernel atomizer ensures that the \texttt{Prelude} kernel uses the same set of resources as the original \texttt{Conv} kernel. Furthermore, the \texttt{Prelude} kernel needs to know the entry point to the \texttt{Conv} kernel. The Kernel Atomizer passes this information to the \texttt{Prelude} kernel in an \texttt{AtomMetadata} struct as seen in Algorithm~\ref{alg:prelude}.

\mypar{Performance Optimizations.} \arch{} continuously monitors the effectiveness of the Kernel Atomizer to enhance performance. First, to avoid the overhead introduced by additional code in the \texttt{Prelude} kernel for kernels with many short threads, \arch{} may disable atomization for such kernels. Additionally, for kernels with a large number of thread blocks, the Kernel Atomizer dynamically adjusts the \texttt{atom\_duration} parameter to control its aggressiveness. This minimizes the performance penalty due to the increased thread block traffic from early-exiting threads.

\subsection{Right-Sizing Hardware Resources}
\label{sec:rightsizing}

\begin{figure}[t]
\centering
\includegraphics[width=\columnwidth]{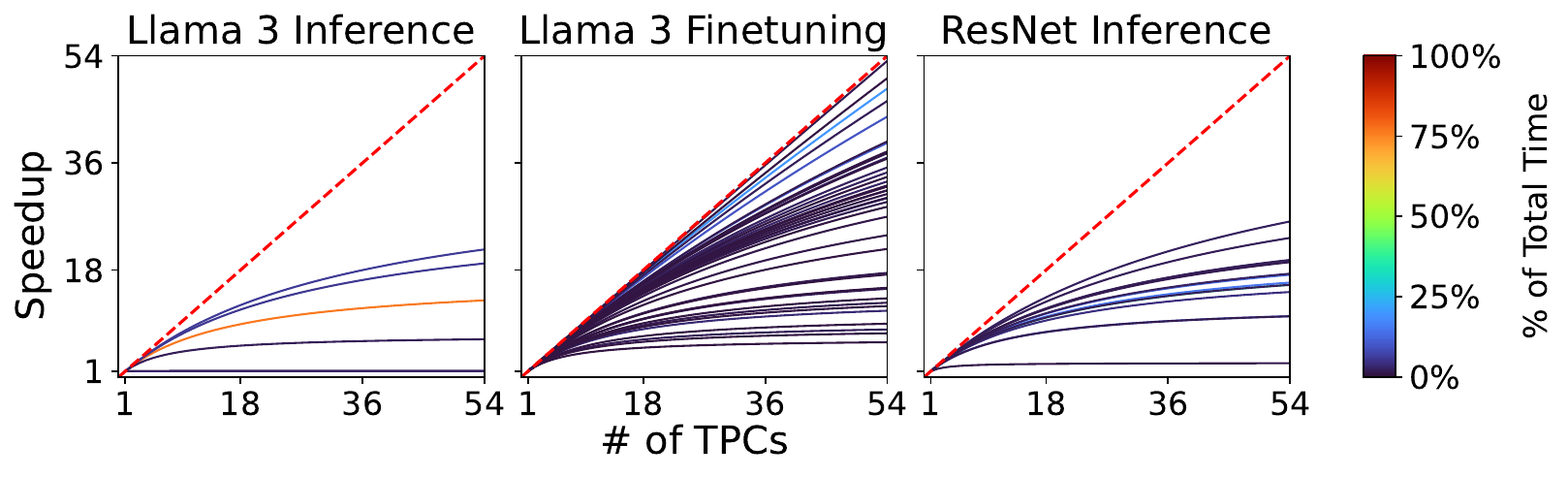}
\vspace{-7mm}
\caption{\arch{}'s interpolated TPC scaling curves.}
\vspace{-4mm}
\label{fig:tpc_scaling}
\Description{}{}
\end{figure}

\arch{}’s ability to schedule at the TPC level unlocks new opportunities for fine-grained GPU right-sizing. Figure~\ref{fig:tpc_scaling} highlights this potential by plotting kernel speedups as a function of allocated TPCs for representative workloads (Section~\ref{sec:methodology}). The selected kernels collectively account for 99\% of total execution time, with color gradients indicating each kernel’s relative contribution. As an example, for Llama inference, general matrix multiplication (GEMM) and multihead attention kernels exhibit diminishing returns, while the kernel responsible for applying the token frequency penalty does not scale. The results show that whole-model right-sizing is suboptimal---there is no single TPC configuration that fits all kernels.
Instead, substantial opportunity lies in right-sizing at the kernel level. First, individual kernels exhibit diverse scaling behaviors: some scale linearly, while others show diminishing returns. Second, the extent to which execution time is distributed across many kernels varies from workload to workload---highlighting the need for adaptive, per-kernel scheduling to fully optimize GPU resource consumption.

\mypar{Modeling Kernel Scaling.} 
\arch{} introduces on-the-fly TPC right-sizing at the granularity of kernels (Figure~\ref{fig:op-timeline}, Step~\circled{6}). The atoms of a given kernel inherit its allocated TPCs, as they exhibit the same scaling behavior as the kernel itself. To this end, \arch{} introduces a model-based approach that interpolates the scaling of individual kernels based on two points: the latencies of a kernel running with all TPCs and just one TPC. It then fits a curve of the form $$l = \frac{m}{t} + b$$ to these points, where $l$ is the predicted latency, $t$ is the corresponding number of TPCs, and $m$ and $b$ are constants. Note that the form of this curve is consistent with Amdahl's law for parallel speedup. Intuitively, $b$ can be thought of as how long it takes for a single one of the kernel's thread blocks to complete on a single SM, and $m$ quantifies the extent to which a kernel can take advantage of parallel processors.

\mypar{Filtering Outliers.}
We find that, in practice, this simple model accurately captures the scaling behavior of most deep learning kernels. However, a small number of outlier kernels---typically those with very short runtimes---deviate from the model, as they fail to benefit from large TPC allocations and are inherently harder to model. To handle these cases, we introduce a \textit{filtering} heuristic based on a kernel’s thread block occupancy. Specifically, we estimate the number of TPCs a kernel can effectively utilize by dividing its total number of thread blocks by the occupancy per TPC---that is, the number of thread blocks a single TPC can execute concurrently. \arch{} already tracks thread blocks per kernel as part of atomization, while occupancy can be queried from the driver API~\cite{occupancyAPI}. This heuristic provides an intuitive upper bound on useful TPC allocations per kernel, helping avoid overprovisioning even for difficult-to-model kernels.

\mypar{Operation.}
When a kernel is submitted to \arch{}, the dispatch thread first applies the filtering heuristic to estimate an upper bound on the number of TPCs the kernel can effectively utilize. If this estimate is lower than the job’s allocated TPCs, the kernel is launched using the estimated bound. Otherwise, the dispatch thread leverages the learned scaling model to determine the minimum number of TPCs that would increase the kernel’s latency by, at most, a multiplicative factor $k$ that we call the \textit{latency slip parameter}. This tunable parameter allows users and administrators to intuitively configure \arch{}, for example, by specifying that up to 10\% performance degradation is acceptable. Overall, \arch{} enables highly efficient fine-grained right-sizing, while its modeling and scaling techniques offer a robust and accurate solution---as we will see in Section~\ref{sec:eval_rightsize}.

\subsection{Transparent Power Management}
\label{sec:dvfs}

\begin{figure}[t]
\centering
\includegraphics[width=\columnwidth]{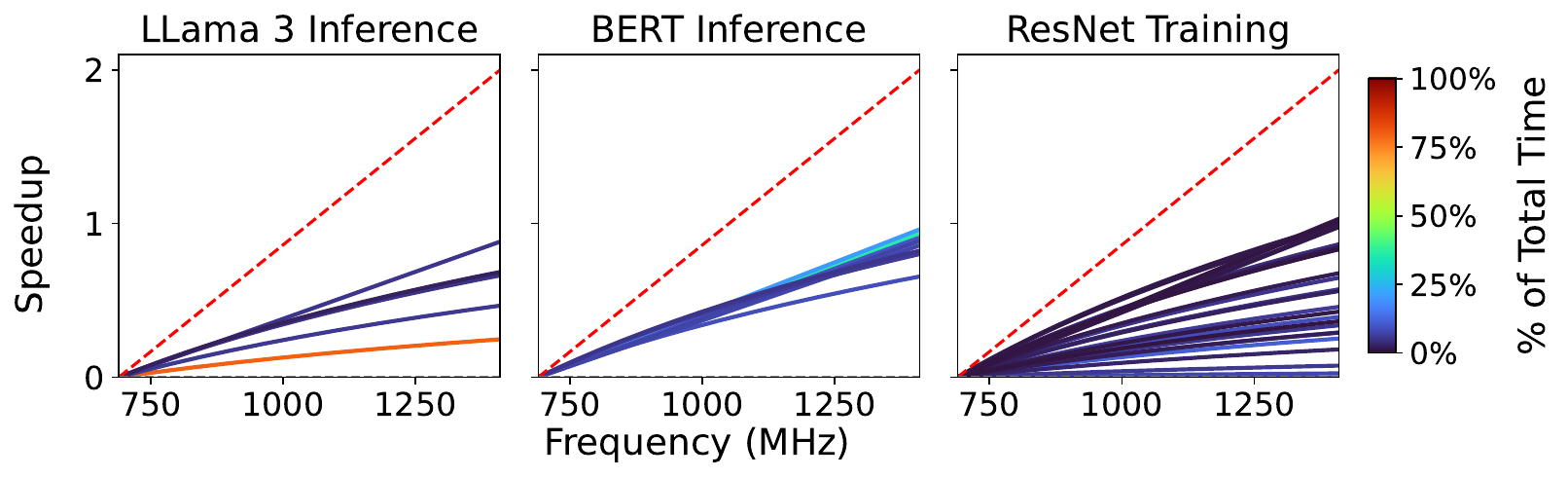}
\vspace{-7mm}
\caption{\arch{}'s interpolated frequency scaling curves.}
\vspace{-4mm}
\label{fig:freq_scaling}
\Description{}{}
\end{figure}

\arch{} is well-positioned to enable transparent and efficient power management via DVFS. Just as right-sizing lets \arch{} adapt resource allocation based on kernel scalability across TPCs, DVFS enables vertical scaling through frequency adjustment. Figure~\ref{fig:freq_scaling} shows how kernels from various workloads respond to frequency scaling. Many exhibit predictable behavior, creating opportunities for energy savings with bounded performance impact.
To achieve efficient DVFS, \arch{} must address two key challenges. First, current GPUs support relatively slow frequency switching ($\approx$50ms). While future architectures may reduce this latency~\cite{blackwellpower}, DVFS remains impractical for models with very short kernels. Thus, \arch{} must consider the cumulative impact of scaling across kernel sequences. Second, although many kernels scale linearly with frequency—enabling significant energy savings—\arch{} must carefully balance these gains against increased latency.

\mypar{Modeling Frequency Scaling.} 
\arch{} introduces a transparent sequence-based kernel frequency scaling model that guides DVFS (Figure~\ref{fig:op-timeline}, Step~\circled{7}). 
Similarly to right-sizing, the atoms of a given kernel inherit its frequency target, as they exhibit the same scaling behavior as the kernel itself. Specifically, each kernel is assigned a weight \textit{w}, the ratio of its total runtime to the cumulative runtime of all the kernels in a particular stream. Then, \arch{} approximates each kernel's relative slowdown as proportional to the fractional drop in frequency based on a first-order Taylor approximation: $$k = \frac{lat(f_{th})}{lat(f_{max})}-1 = s\cdot(\frac{f_{max}}{f_{th}}-1)$$ where $lat(f)$ is the kernel's latency at frequency $f$. Specifically, $f_{max}$ is the maximum frequency, and $f_{th}$ is one of the device's supported frequencies. 
Each kernel's sensitivity is $$s = \frac{k}{\frac{f_{max}}{f_{th}}-1}$$ and the aggregate sensitivity S across all kernels is equal to $\sum{w*s}$. Similarly, the total slowdown is equal to $$S \cdot (\frac{f_{max}}{f_{final}} - 1) \leq k$$ and thus the final frequency that \arch{} assigns to the workload is $f_{final} = \frac{f_{max}}{1 + \frac{k}{S}}$. Intuitively, compute-bound kernels whose slowdown scales linearly with frequency reduction skew the final frequency closer to the maximum according to their sensitivity, while memory-bound kernels whose slowdown is frequency‑insensitive skew the final frequency to lower levels depending on their weight. 

\mypar{Operation.} Similar to right-sizing, \arch{} uses a multiplicative factor $k$, the \textit{latency slip parameter}, to guide DVFS decisions. At runtime, this parameter is used to evaluate the scaling model and select a target frequency. Due to the high latency of switching, \arch{} adopts a conservative strategy and extends its learning period to avoid unnecessary transitions.
Initially, \arch{} collects per-kernel metadata at maximum frequency, forcing unseen kernels to run at max frequency. On first appearance, a kernel is assumed to scale linearly, and its frequency is reduced based on the configured $k$. Depending on the observed performance, \arch{} either further lowers the frequency or stops after confirming linear behavior. Over time, it fits the collected data to the scaling model described earlier, enabling more informed and efficient DVFS decisions.

\subsection{Online Latency Prediction}
\label{sec:learning}

The latency prediction module learns the execution time of kernels, enabling the optimizations carried out by all of \arch{}'s components. In particular, it enhances TPC Stealing by estimating the duration of outstanding tasks and guides the number of atoms the Kernel Atomizer splits each kernel into. It further assists right-sizing and DVFS by providing the latencies that are used to calculate speedups based on TPCs and frequency scaling. This obviates the need for extensive offline profiling, which is impractical for a transparent OS.

Latency prediction operates separately for independent launch queues, allowing \arch{} to dynamically adapt to the behavior of different applications. During execution, the module records kernel latencies and uses this data to refine its predictions. Each kernel's latency varies based on the allocated TPCs, the GPU frequency, and the granularity at which it is atomized; therefore, the prediction module must monitor these conditions to produce accurate estimates. In the case where such metadata are not available for a specific atom, the prediction module is conservative, assuming optimal linear scaling. For instance, if an atom was previously executed with a TPC allocation of 100\%, it fits a linear trend to estimate the duration when given half of the GPU.

One pitfall in achieving accurate kernel latency prediction is assuming that a given kernel function always has the same latency. In actuality, the duration can depend on the kernel's launch parameters and input arguments. For instance, a single \texttt{Conv} kernel function can be used multiple times across model layers that have varying tensor sizes. This necessitates that the latency prediction module track operators rather than kernel functions themselves. 

By recording explicit synchronization events from the application, we can determine the start and end of a batch. We associate kernel launches with an ordinal index $k$, referring to the $k^{\text{th}}$ kernel after the start of a batch. This uniquely identifies operator nodes in the model's data flow graph (DFG), despite \arch{} lacking explicit access to this high-level information. This additional ordinal index is sufficient to identify model operators and make accurate latency predictions. 

\section{Implementation}
\label{sec:implementation}

We implement a prototype of \arch{} targeting NVIDIA GPUs in $\sim$5000 lines of Rust, excluding macro-generated code for interposing the entire CUDA Driver API. Our prototype supports applications running natively or in containers. To enable concurrent execution across GPU contexts, we build on top of MPS.
We defer some low-level details of the implementation to a separate technical report.

\mypar{Interposition Architecture.} \arch{} is fully transparent to applications, supporting the diverse ML ecosystem and full GPU stack. A key implementation decision is abstracting the CUDA Driver API, the lowest common denominator across the stack. Instead of accessing the driver directly, applications interact with \arch{}, which preserves CUDA call semantics. This abstraction ensures generality and transparency at the OS level. \arch{} seamlessly supports unmodified ML applications using frameworks and libraries like PyTorch, TensorFlow, JAX, TensorRT, and closed-source libraries like cuDNN.
Beyond transparency, \arch{} avoids the complexity of interposing across the CUDA Runtime and other libraries that eventually call into the driver. Instead, it implements a small subset of CUDA APIs (e.g., \texttt{cuLaunchKernel}), while the rest are auto-generated via our toolchain. The \arch{} library also eliminates complex data marshaling across address spaces, unlike prior CUDA API interposition systems~\cite{ava}. This approach enables rapid support for new CUDA versions with minimal effort, enhancing long-term OS maintainability.

\mypar{Isolation and Faults.} In \arch{}, applications run in separate address spaces and cannot access each other’s memory. Illegal accesses lead to termination of the offending application. To handle other faults, \arch{} enables graceful termination for common errors~\cite{Muxflow} by intercepting signals and terminating the application without affecting other contexts.
\section{Experimental Setup and Methodology}
\label{sec:methodology}

\mypar{Testbed.} Experiments were conducted on a \texttt{1x A100 (SXM4)} Lambda Labs GPU instance with 30 CPU cores and 216~GB of host memory. The A100 GPU has 108 SMs with 40 GB of memory. The server was configured with Ubuntu 22.04, CUDA 12.8, Rust 1.83.0-nightly, Python 3.10, PyTorch 2.3, TensorRT 10.7, TensorRT-LLM 0.16.0, and Triton 24.12.

\mypar{Baselines.} 
We compare \arch{} with all four NVIDIA GPU sharing methods: \textit{Time slicing}, \textit{MPS}, stream \textit{Priority}, and \textit{MIG}. We further compare against SOTA prior work across the spectrum of transparent solutions \textit{TGS}~\cite{tgs}, application modifications \textit{REEF}~\cite{reef}, and both application modifications and offline profiling \textit{Orion}~\cite{orion}. We used the open-source TGS directly but had to re-implement Orion and REEF using our own interposition infrastructure, since the available code was tied to specific CUDA drivers and software stacks. We extend REEF and Orion to handle multiple HP apps in a straightforward manner. For REEF, BE kernels are not launched if \textit{any} HP app is running. For Orion, BE kernels are not launched if they contend with \textit{any} HP kernel.

\begin{table}[]
\footnotesize
\begin{tabular}{c c c c}
\hline
Model & Mem. (GiB) & Batch Size & Latency (ms) \\
\hline
VGG-19~\cite{simonyan2015deepconvolutionalnetworkslargescale} & 17.4 & 120 & 291  \\
ResNet-50~\cite{he2015deepresiduallearningimage} & 18.4 & 184 & 281 \\
MobileNetV2~\cite{sandler2018mobilenetv2} & 18.4 & 216 & 254 \\
DLRM~\cite{naumov2019deeplearningrecommendationmodel} & 6.7 & 32768 & 74 \\
BERT-Large~\cite{devlin2019bertpretrainingdeepbidirectional} & 17.3 & 20 & 159  \\
Llama 3 Finetuning & 32.0 & 4 & 690 \\
\hline
\end{tabular}
\caption{Training model parameters.}
\label{table:train_models}
\vspace{-10mm}
\end{table}

\mypar{Models and Configurations.} All high priority inference tasks run on NVIDIA's Triton Inference Server with dynamic batching~\cite{Triton}. RetinaNet runs on ONNX Runtime while the other served models run on NVIDIA's TensorRT and TensorRT-LLM backends.
We choose three representative vision models (RetinaNet~\cite{lin2018focallossdenseobject}, YOLOv4~\cite{bochkovskiy2020yolov4optimalspeedaccuracy}, and ResNet-50 v1.5~\cite{he2015deepresiduallearningimage}) and three language models (Llama 3 8B~\cite{dubey2024llama3herdmodels}, GPT-J 6B~\cite{gpt-j}, and BERT-Large~\cite{devlin2019bertpretrainingdeepbidirectional}) as inference workloads. For large language models, we use a Microsoft Azure trace~\cite{stojkovic2024}. For the best effort training tasks, we use three vision models, ResNet-50, MobileNetV2, VGG-19, a deep learning recommendation model (DLRM), a language model BERT-Large, and LLM fine-tuning with Llama 3 as listed. The training batch size is adjusted to use at most half of the GPU DRAM to keep all models in memory when stacking. The best effort training task runs continuously. More details are shown in Table~\ref{table:train_models}.

\section{Evaluation}
\label{sec:eval}

Our evaluation answers the following questions:
\begin{enumerate}
    \item Does \arch{} improve performance for different multi-tenancy environments and SOTA prior works?
    \item What are the capacity savings due to \arch{}'s hardware right-sizing?
    \item What are the energy savings of \arch{}'s DVFS?
    \item How do different \arch{} features affect performance?
\end{enumerate}

\subsection{Performance in Multitenant Environments}
In the following experiments, we disable right-sizing and power management features of \arch{} to provide an apples-to-apples comparison to other systems in terms of scheduling efficiency alone. We evaluate these features afterwards.

\begin{table}
\footnotesize
\begin{tabular}{c c c c}
\hline
Model & Framework & Load (rps) & Constraint (ms) \\
\hline
ResNet~\cite{he2015deepresiduallearningimage} & TensorRT & 1000 & 15 \\
RetinaNet~\cite{lin2018focallossdenseobject} & ONNX Runtime & 9 & 100 \\
Llama 3~\cite{dubey2024llama3herdmodels} & TensorRT-LLM & 0.5 & 2000 \\
GPT-J~\cite{gpt-j} & TensorRT-LLM & 0.5 & 2000 \\
BERT~\cite{devlin2019bertpretrainingdeepbidirectional} & TensorRT & 30 & 130 \\
\hline
\end{tabular}
\caption{Inference services for inference-only multitenancy.}
\label{table:inf_inf}
\end{table}

\mypar{Inference-only Multitenancy.} We evaluate \arch{} in a multitenant environment with three inference applications: one best-effort (BE) and two high-priority (HP). The first HP app, \textit{HP A}, has a latency-oriented SLO: percentage of requests executed within a latency constraint. The second, \textit{HP B}, has a throughput-oriented SLO: attained throughput as a percentage of the case where it executes alone. These vary according to the model (Table~\ref{table:inf_inf}).

The BE and HP B models are chosen from Llama 3, GPT-J, and BERT. For HP A, we add ResNet and RetinaNet. We use latency constraints from the MLPerf Datacenter inference benchmark~\cite{mlperf} (Table~\ref{table:inf_inf}). 
We run all possible combinations. HP apps follow Poisson load and run on the Triton inference server, while BE apps execute in a closed loop. Latencies for all models, including LLMs, are measured end-to-end.

We compare \arch{} against all configurations. For systems that support partitioning, HP A and HP B are isolated on partitions of 75\% and 25\%, respectively. MIG's limited partitioning configurations cannot support a 25\%-75\% split, so we use a 3/7-4/7 split instead. MIG and Limits cannot support a BE app, but only apps with provisioned resources; therefore, the BE does not run on these systems. There is no way to isolate multiple latency-sensitive applications on systems like Priority, REEF, TGS, and Orion. For these, we set both of the HP apps to high priority and the BE to low priority.

\begin{figure}[t]
\centering
\includegraphics[width=0.65\linewidth]{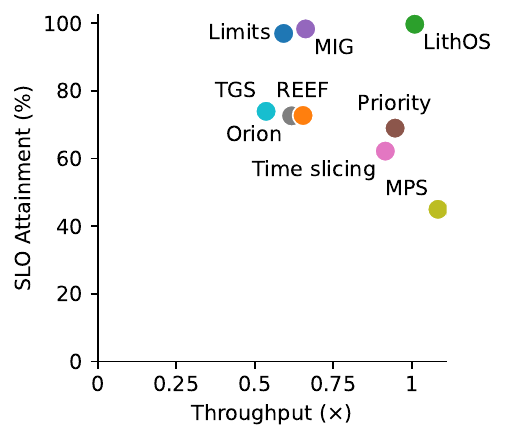}
\vspace{-4mm}
\caption{SLO attainment and throughput by system.}
\vspace{-2mm}
\label{fig:inf_inf/trade-off}
\end{figure}

Figure~\ref{fig:inf_inf/trade-off} compares all systems across two dimensions: SLO attainment and throughput. 
``SLO'' of 100\% means both HPs reach 100\% attainment.
``Throughput'' of 1 means that the throughput achieved is as much as if any of the apps had the entire device.
Unsurprisingly, MPS sets the bar for throughput. MPS's fine-grained, intra-SM stacking ensures device resources are maximally utilized, and it allows more throughput when stacking than any application could have alone; hence, it achieves a throughput of 1.08. MPS's throughput comes at the cost of SLO attainment, at 42\%. MIG and thread limits both successfully meet SLOs. This is expected, as each system minimizes interference by devoting resources to individual apps. However, the partitions are not fully utilized without a BE app. As a result, aggregate throughput drops to 0.66 and 0.59 for thread limits and MIG, respectively. Without isolating HP apps, priority-only systems cannot attain SLOs, with TGS leading at 72\%. \arch{} provides the best of both worlds, as it provides spatial isolation like MIG with an SLO attainment of 100\% and a throughput of 1.

\begin{figure}[t]
\centering
\includegraphics[width=\linewidth]{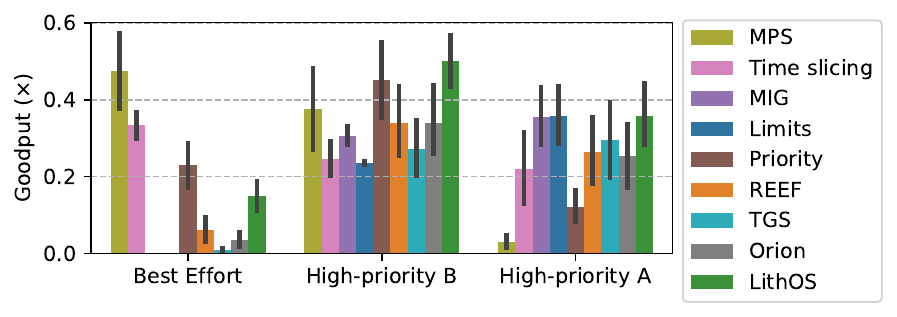}
\vspace{-8mm}
\caption{Inference-only multitenancy: goodput by app.}
\vspace{-2mm}
\label{fig:inf_inf/goodput}
\end{figure}

Where do the benefits of \arch{} come from? Figure~\ref{fig:inf_inf/goodput} shows \arch{} consistently leading in goodput (throughput excluding HP A requests that violate the SLO constraint) for the HP apps while still allowing significant (0.15) BE throughput. While the partitioning systems match \arch{} in HP A goodput, they lack in HP B goodput: MIG at 0.31 vs. \arch{} at 0.50. Partitioning schemes cannot support any BE throughput, while \arch{} allows HP apps to steal unused resources from one another and further support BE throughput. No SotA system can perform effectively across all requirements. Specifically, REEF and Orion underperform in latency-sensitive goodput and TGS in throughput. Only \arch{} provides the best HP goodput while sustaining high BE throughput. 

\begin{figure*}[t]
\centering
\includegraphics[width=\textwidth]{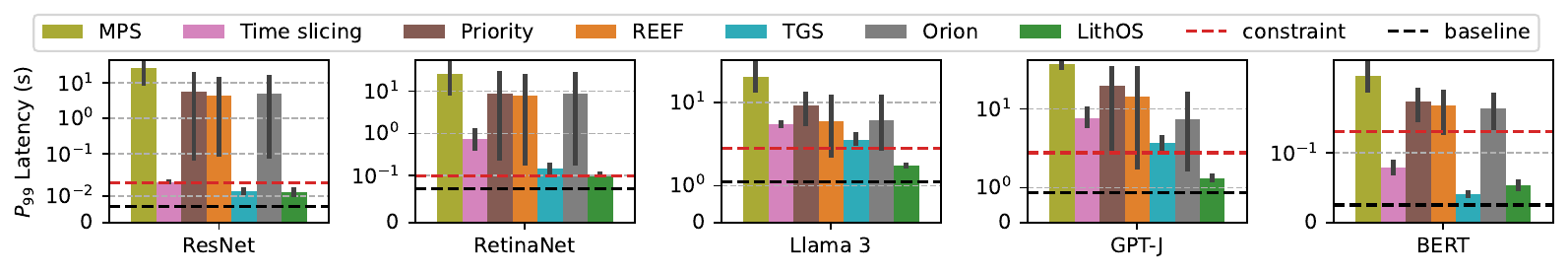}
\vspace{-8mm}
\caption{Inference stacking multitenancy: \textit{HP A} tail latencies by model.}
\vspace{-2mm}
\label{fig:inf_inf/latency}
\end{figure*}

Diving deeper, we next look into the latencies of the HP A app in Figure~\ref{fig:inf_inf/latency}.
The figure shows the $P_{99}$ latencies for each model averaged across all combinations. Latencies diverge in many cases, with only \arch{} and the partitioning systems limiting latencies to the constraints. MPS is the worst-performing regarding latencies; \arch{}'s latencies are 13\x{} better. \arch{} reduces latencies by 12\x{} compared to Orion. This is expected as Orion cannot handle multiple HP apps. TGS limits latencies much more effectively than Orion and REEF, but \arch{} still improves over it by 3\x{}. Overall, \arch{} provides a robust solution for inference stacking. 

\begin{figure*}[t]
\centering
\includegraphics[width=\textwidth]{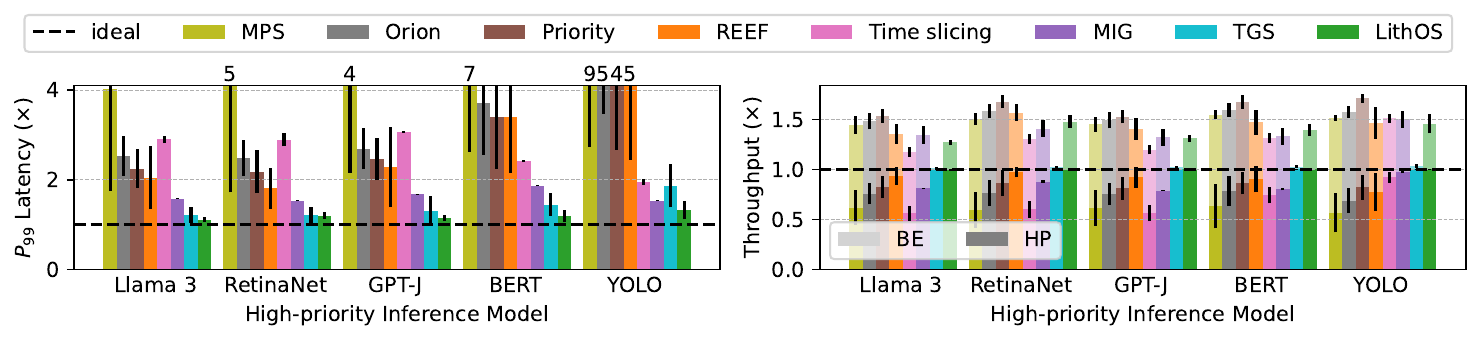}
\vspace{-8mm}
\caption{Hybrid multitenancy: (a) $P_{99}$ service latency and (b) aggregate throughput.}
\vspace{-2mm}
\label{fig:inf-train}
\Description{}{}
\end{figure*}

\mypar{Hybrid Inference/Training Multitenancy.} In this experiment, we stack an HP that has a latency-oriented SLO with a training BE app. Similar to the inference-stacking experiment, resources unused by the sensitive inference app should be donated to the best-effort training job. At the same time, service latency must not increase. We choose the inference model from the set, Llama 3 8B, GPT-J 6B, BERT-Large, RetinaNet, and YOLOv4. We choose the training model from those listed in Table~\ref{table:train_models}. We run all model combinations, and our client creates Poisson loads. Load parameters are chosen to keep GPU utilization around 80\% for the HP app.

Figure~\ref{fig:inf-train} shows the $P_{99}$ HP latency and aggregate throughput, averaged across all training models. HP throughput is normalized to the load before being added to the BE throughput, normalized to the case where the BE model runs alone on the device. 
Latencies are also normalized to the case where the HP runs alone on the device.
MPS yields latencies 5.83\x{} the ideal case, and its service throughput is the lowest at 60\%. Time slicing fares better as it enables the long-running kernels of the best-effort models to be preempted, guaranteeing the service approximately 50\% of the GPU time.  
MIG performs similarly to time slicing by allocating 50\% of the GPU to the service spatially rather than temporally. However, both methods fail to sustain peak HP throughput. 
Stream priority also falls short, leading to a 2.89\x{} increase in service latency and service throughput as low as 68\%.

Both TGS and REEF also struggle to maintain low service latencies. TGS has an average inference latency of 1.41\x{} the ideal, and REEF 2.89\x{}. TGS's poor performance stems from its adaptive rate control mechanism, which assumes a constant work arrival rate. This assumption is invalid for inference services, which have unpredictable load patterns. REEF fails to sufficiently throttle the training model, allowing tail latencies to reach 8.93\x{}.
In contrast, \arch{} maintains a tail latency within 20\% of the ideal. On average, this is a 2.34$\times$ and \infTrainTailTGS{}\x{} over REEF and TGS, respectively. Compared to the native MPS solution, \arch{} reduces latency by up to 13.54\x{} and \infTrainTailMPS{}\x{} on average. \arch{} maintains service throughput within 1\% of load in the worst case. \arch{} improves training throughput by an average of 34\x{} and aggregate throughput by \infTrainTptTGS{}\x{} vs. TGS. In total, \arch{} improves aggregate throughput 1.23\x{}--1.57\x{} with an average of 1.38\x{}.

\subsection{Kernel-SM Right-Sizing}
\label{sec:eval_rightsize}

\begin{figure}[t]
\centering
\includegraphics[width=0.94\columnwidth]{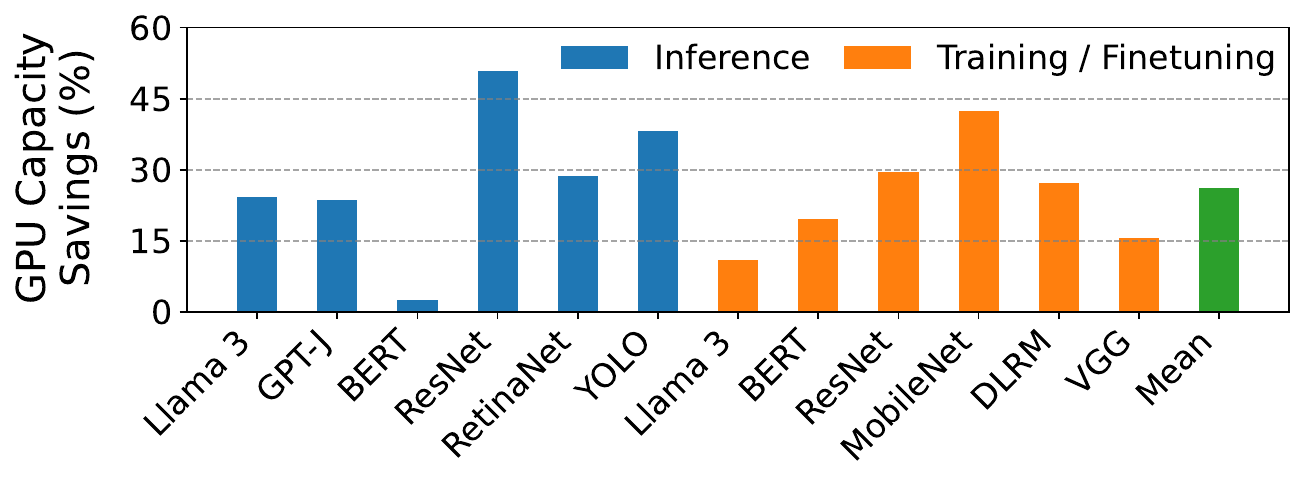}
\vspace{-4mm}
\caption{Hardware right-sizing GPU capacity savings.}
\label{fig:capacity_savings}
\vspace{-4mm}
\Description{}{}
\end{figure}

\mypar{Capacity Savings.} Figure~\ref{fig:capacity_savings} shows the capacity savings due to right-sizing with \arch{}. We compute savings by comparing the time-weighted average of TPC utilization before and after right-sizing. \arch{} provides excellent savings of up to 51\%, and a mean of \rightsizingAVG$\%$ across all workloads. We expect that in future GPU architectures with an increased number of TPCs, the fine-grained right-sizing approach of \arch{} will provide even more aggressive saving potential. 

\mypar{Latency and Throughput Cost.} With a latency slip parameter of 1.1, the performance cost of right-sizing in terms of $P_{99}$ and throughput is modest. The mean increase in $P_{99}$ and decrease in throughput are both \rightsizingPerfAVG\%. Our latency slip parameter is conservative because not all of the end-to-end execution time of each inference or training iteration is spent inside a GPU kernel; this does not impede tuning in practice.

\mypar{Accuracy.} To quantify the accuracy of our prediction technique, we compute the kernel-execution-time weighted average of the $R^2$ values for the curves we fit (i.e., for kernels where the possible TPCs value exceeds the threshold). Across all of the evaluated workloads, the average $R^2$ values range from 0.92 (Llama finetuning) to 0.99 (RetinaNet inference), indicating that our technique is highly accurate.

\subsection{Kernel-Dependent DVFS}

\begin{figure}[t]
\centering
\includegraphics[width=\columnwidth]{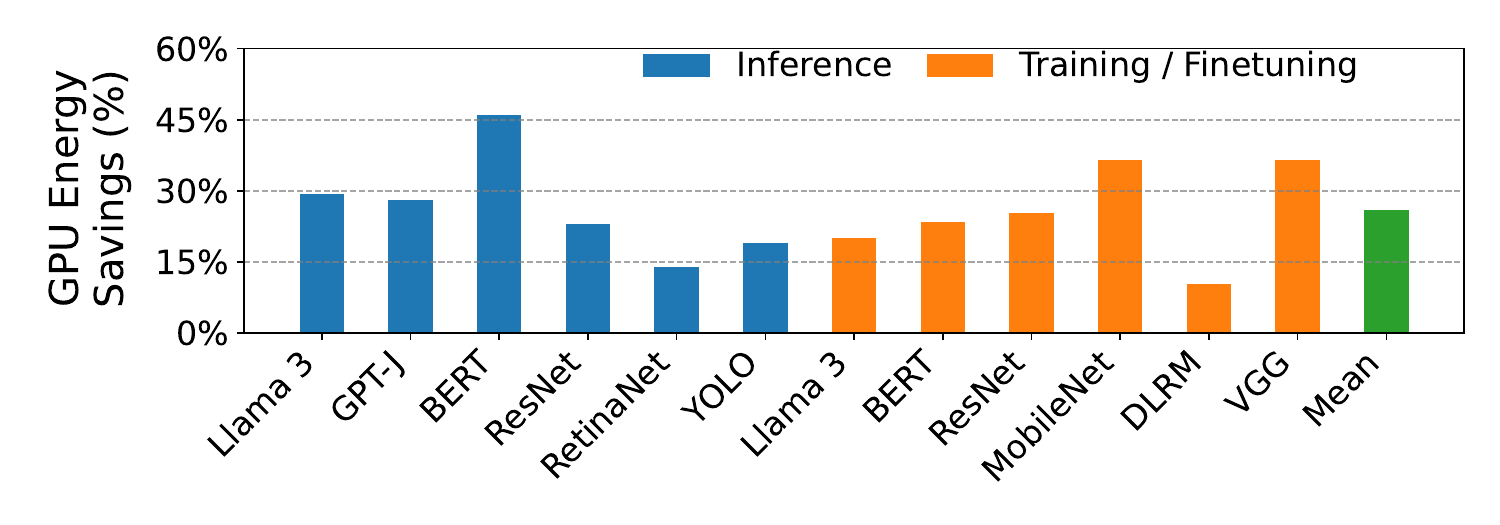}
\vspace{-8mm}
\caption{Power management GPU energy savings.}
\vspace{-2mm}
\label{fig:energy_savings}
\Description{}{}
\end{figure}

\mypar{Energy Savings.} Figure~\ref{fig:energy_savings} shows the energy savings of \arch{}'s DVFS mechanism across different inference and training workloads.
We define energy savings by recording the difference between executing the workload at maximum frequency, and under \arch{}'s DVFS policy. 
\arch{} provides significant energy savings of up to 46\%, and a mean of 26\% across all workloads without offline profiling requirements.  

\mypar{Performance Cost.} The slip parameter for this experiment was set at 1.1, and the mean increase in $P_{99}$ latency is 7\%.
The minimal increase in $P_{99}$ latency demonstrates that \arch{}'s DVFS policy is inherently conservative. It respects latency constraints across workloads while transparently providing substantial energy savings. Finer-grained frequency control could unlock additional energy savings.

\subsection{Ablation and Case Studies}

\begin{figure}[t]
\centering
\includegraphics[width=0.9\linewidth]{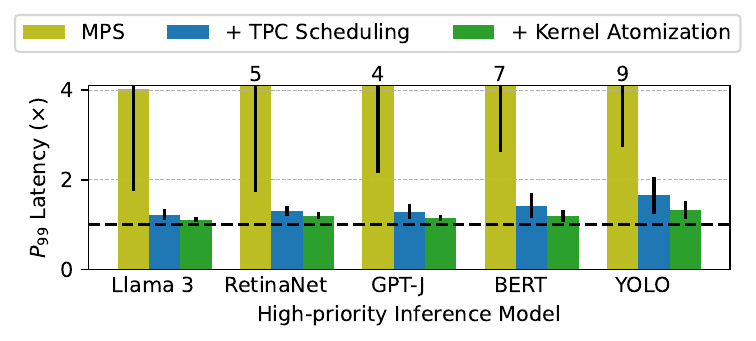}
\vspace{-5mm}
\caption{Breakdown of \arch{} features for inf-train.}
\vspace{-2mm}
\label{fig:ablation_inf_train}
\Description{}{}
\end{figure}

\mypar{Multi-tenancy Breakdown.} Figure~\ref{fig:ablation_inf_train} presents a performance analysis for inference-training as explored in Figure~\ref{fig:inf-train}. Enabling the TPC scheduler improves HP tail latencies to 1.38\x{} ideal by throttling BE work, while maintaining ideal HP throughput.
Kernel Atomization offers additional gains, reducing tail latencies to an average of 1.19\x{} and up to 1.55\x{}, by splitting long BE kernels and improving TPC Stealing. Because of space limitations, we plot only latencies. Kernel Atomization introduces a 10\% throughput overhead, as \arch{} prioritizes HP workloads by reducing BE throughput. Overall, each of \arch{}'s features plays a crucial role in optimizing end-to-end performance.

\begin{figure}[t]
\centering
\includegraphics[width=0.97\linewidth]{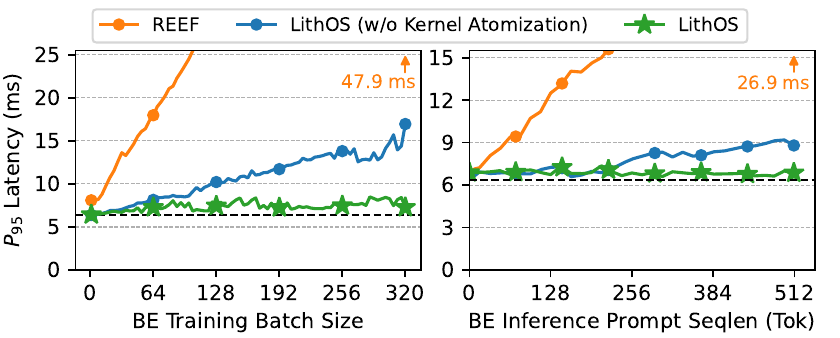}
\vspace{-3mm}
\caption{ $P_{95}$ latency of HP inference collocated with varied (a) batch sizes training and (b) sequence lengths inference.}
\vspace{-3mm}
\label{fig:var-batch}
\Description{}{}
\end{figure}

\mypar{Kernel Atomization.} To highlight the challenges of scheduling long-running kernels, we collocate an HP BERT inference workload with either a BE VGG training or a BE Llama 3 inference. 
In Figure~\ref{fig:var-batch}, we vary (a) the batch size of the BE training job and (b) the sequence length of the BE inference job and measure the $P_{95}$ latency of the HP inference job. \arch{} outperforms REEF by 6.5$\times$ and 3.9$\times$ in (a) and (b), respectively. Unlike REEF, which simply throttles BE work, \arch{} accounts for kernel durations, which can vary significantly.
To understand the impact of Kernel Atomization, we further evaluate \arch{} with Kernel Atomization disabled.
Kernel Atomization provides an improvement of 2\x{} and 1.3\x{} in (a) and (b), respectively. As described in  Figure~\ref{fig:latency_vs_bs}, kernel durations grow with training batch size and inference input sequence length. As Kernel Atomization allows \arch{} to schedule at thread block granularity, HoL blocking is minimized. Consequently, the HP tail latency for the full \arch{} system is within 14\% (or 1\ms{}) or 7\% (or 0.45\ms{}) of ideal for even the largest batch size or sequence length, respectively.

\mypar{Latency Prediction Module.}
Next, we evaluate the accuracy of the latency prediction module of \arch{} that enhances the TPC Scheduler and the Kernel Atomizer. Specifically, we record the predicted atom latencies and compare them with the corresponding recorded CUDA events. We consider absolute errors greater than 50\us{} to be mispredictions.
Overall, we find very low misprediction rates of just 0.9\% and 0.38\% for the HP workloads in inference-inference and inference-training environments, respectively. Additionally, the prediction error tails are small with $P_{99}$ values of 49\us{} and 31\us{}.
Misprediction rates for the BE workloads are higher at 14\% and 11\% for inference-inference and inference-training, respectively. This is acceptable as BE work is frequently preempted by HP work and has lower priority for GPU resources.
\section{Conclusion}
This paper introduces \arch{}, a first step towards an operating system for efficient machine learning on GPUs. \arch{} operates transparently to the entire ML stack.
Through mechanisms like TPC Scheduling, Kernel Atomization, hardware right-sizing, and power management,
\arch{} significantly improves GPU efficiency while laying the foundation for future OS research on GPUs.

\section{Acknowledgments}
We would like to thank the members of the Computer Architecture and Operating Systems (\href{https://www.cs.cmu.edu/~caos/}{CAOS}) group at the Computer Science Department at Carnegie Mellon University for their feedback on this work. This research was funded by a Meta AI Hardware/Software Co-design Faculty award and NSF grants CNS 2239311 and CCF 2217016.

\bibliographystyle{ACM-Reference-Format}
\bibliography{references,models}

\end{document}